\documentclass[twocolumn,showpacs,preprintnumbers,amsmath,amssymb,noshowpacs, prl, superscriptaddress]{revtex4-1}
\usepackage{graphicx}
\usepackage{dcolumn}
\usepackage{mathrsfs}
\usepackage{bm}
\usepackage{color}

\begin{document}

\preprint{}

\title{Subnatural-linewidth biphotons from a Doppler-broadened hot atomic vapor cell}


\author{Chi Shu}
\affiliation{Department of Physics, The Hong Kong University of Science and Technology, Clear Water Bay, Kowloon, Hong Kong, China}

\author{Peng Chen}
\affiliation{Department of Physics, The Hong Kong University of Science and Technology, Clear Water Bay, Kowloon, Hong Kong, China}

\author{Tsz Kiu Aaron Chow}
\affiliation{Department of Physics, The Hong Kong University of Science and Technology, Clear Water Bay, Kowloon, Hong Kong, China}

\author{Lingbang  Zhu}
\affiliation{Department of Physics, The Hong Kong University of Science and Technology, Clear Water Bay, Kowloon, Hong Kong, China}

\author{Yanhong Xiao}
\affiliation{Department of Physics, State Key Laboratory of Surface Physics, Key Laboratory of Micro and Nano Photonic Structures, Fudan University, Shanghai 200433, China}

\author{M. M. T. Loy}
\affiliation{Department of Physics, The Hong Kong University of Science and Technology, Clear Water Bay, Kowloon, Hong Kong, China}

\author{Shengwang Du}\email{Corresponding author: dusw@ust.hk}
\affiliation{Department of Physics, The Hong Kong University of Science and Technology, Clear Water Bay, Kowloon, Hong Kong, China}

\date{\today}

\maketitle

\begin{figure*}
\includegraphics[width=17cm]{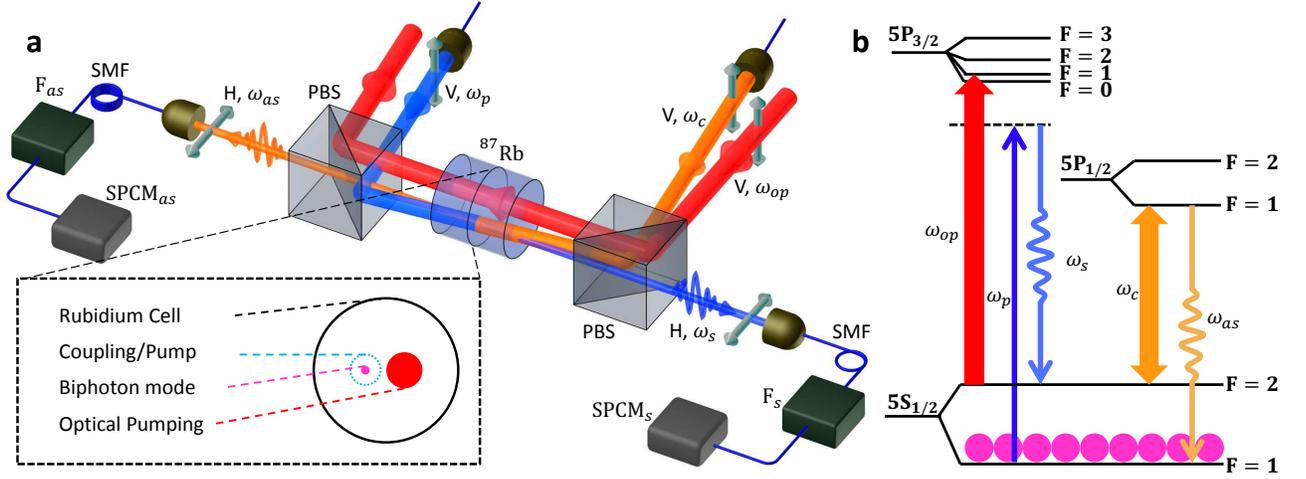}
\caption{\label{fig:schematic} \textbf{Generating narrowband biphotons from a hot $^{87}$Rb vapor cell.} \textbf{a}, Experimental setup. In presence of counter-propagating vertically (V) polarized pump (780nm, $\omega_p$) and coupling (795nm, $\omega_c$) laser beams, horizontally polarized (H) Stokes (780nm, $\omega_s$) and anti-Stokes (795nm, $\omega_{as}$) photon pairs are spontaneously generated and coupled into two opposing single-mode fibers (SMF). Then they pass through optical frequency filters ($F_{s}$, and $F_{as}$), and detected by two single-photon counting modules (SPCM$_s$ and SPCM$_{as}$).The fiber-fiber coupling efficiency and SPCM detection efficiency are 80$\%$ and 50$\%$ respectively. Each optical frequency filter compose of a wide-band line filter and a narrowband etalon Fabry-Perot cavity filter. The etalon filters have free spectrum range FSR=13.6 GHz. The bandwidth, transmission efficiency and the extinction ratio of the frequency filters are 350MHz, 80$\%$ and 60 dB for $F_s$; and 80MHz, 30$\%$ and 40dB for $F_{as}$.  In the inset, we make a zoom-in on the transverse cross section of cell to clearly show the beam profiles of all incident lasers and the biphoton mode. The biphoton mode has a waist diameter ($1/e^2$) of 250 $\mu$m focused in the middle of the 0.5-inch-long cell. The collimated coupling and pump laser beams are counter propagating and have the same $1/e^2$ beam diameter of 1.4 mm.  The optical pumping beam has a $1/e^2$ diameter of 2 mm and does not overlap with the SFWM volume enclosed by the pump and coupling beams.  The cell inner diameter is 10 mm. \textbf{b}, The relevant $^{87}$Rb atomic energy level diagram for backward SFWM and optical pumping. The strong optical-pumping laser is used to optically pump the atoms from the level $|5S_{1/2}, F=2\rangle$ to $|5S_{1/2}, F=1\rangle$ to suppress the on-resonance Raman scattering of the coupling beam.}
\end{figure*}

\noindent\textbf{Entangled photon pairs, \textit{biphotons}, have been the bench-mark tool for experimental quantum optics \cite{Quantuminformation}. The quantum-network protocols based on photon-atom interfaces have stimulated a great demand for single photons with bandwidth comparable to or narrower than the atomic natural linewidth \cite{QuantumNetwork}. In the past decade, laser-cooled atoms has been often used for producing such biphotons, but the apparatus is too large and complicated for engineering \cite{Subnatural, KurtsieferPRL201, DuOptica2013}. Here we report the efficient generation of subnatural-linewidth ($<$ 6 MHz) biphotons from a Doppler-broadened (530 MHz) hot atomic vapor cell. We use on-resonance spontaneous four-wave mixing (SFWM) in a hot paraffin-coated $^{87}$Rb vapor cell at 63 $^\circ$C to produce biphotons with controllable bandwidth (1.9 - 3.2 MHz) and coherence time (47 - 94 ns). Our backward phase-matching scheme with optical pumping is the key to minimize the generation of uncorrelated photons from resonance fluorescence. The result paves the way toward miniature narrowband biphoton source based on atomic cells.}

Time-frequency entangled photon pairs are characterized by the bandwidth of their two-photon joint spectrum. In earlier days, spontaneous parametric down conversion using nonlinear crystals \cite{HarrisPRL1967, WeinbergPRL1970} and four-wave mixing in optical fibers \cite{RarityOE2005} were standard methods for producing biphotons. However, biphotons produced directly using these methods have very wide bandwidth ( $>$ THz) and short coherence time ($<$ ps), which make them extremely difficult for practical implementation of photonic quantum information processing and quantum communication. For example, atomic quantum memory-repeater based long-distance quantum communication network requires single photon bandwidth to be narrower than the atomic natural linewidth for efficiently absorbing and storing the photonic quantum states in atomic media \cite{GisinPRL2007}. Another example is the quantum state teleportation whose distance is limited by the path length difference of the interacting photons that is required to be stabilized within their temporal coherence time \cite{GisinPRL2004}. This is the reason why most early experiments required the use of optical filters to narrow down the photon bandwidth, but with reduced generation efficiency and photon brightness. To solve this problem, many researches have focused on narrowing down the paired photon bandwidth ($\sim$10 MHz) and improving spectral brightness by putting the nonlinear crystal inside an optical cavity \cite{LuPRL1999, PanPRL2008}. However generation of biphotons from solid-state materials with bandwidth narrower than the natural linewidth (6 MHz) of rubidium D1/D2 lines has not been demonstrated.

So far, subnatural-linewidth biphotons have only been produced from SFWM in cold atoms (10-100 $\mu$K) assisted with electromagnetically induced transparency (EIT) \cite{Subnatural, DuOptica2013, ChenSR2015} or cavity \cite{VuleticScience2006}. However, cold-atom systems are not freely accessible to most quantum optics researchers who have no expert knowledge in laser cooling and trapping. A cold atom apparatus is not only expensive, but also large in its size and complicated in its vacuum-optical-electronic-mechanical configuration. Moreover, operating cold atoms for producing paired photons requires a complex timing control with a low duty cycle \cite{DuRSI2012} and the SFWM process in cold atoms cannot be run continuously. Therefore, using cold atoms maybe ideal for fundamental research, but not suitable for practical applications and engineering.

If a hot or room-temperature atomic vapor cell can be used as an alternative atomic source to produce narrow-band biphotons, the system size and operation can be dramatically simplified and the cost will be significantly reduced. However, the use of hot atomic vapor cell for producing narrow-band biphotons has not been as successful as those with cold atoms. Many researches of multi-wave nonlinear mixing have been carried out with hot atomic vapor cells \cite{LettOL2007, HowellNP2009, XiaoPRL2009}, but only very limited few worked at the single photon level. In an early demonstration in 2005, Lukin \textit{et al.} generated non-classical correlated light pulses from a room-temperature $^{87}$Rb atomic vapor cell with “writing-reading” pulse operation \cite{LukinNature2005}, but these photons are not time-frequency entangled and the photon number in each pulse is barely below the two-photon threshold. In this work, we focus on paired photon generation with time-frequency entanglement and the continuous-wave (cw) operation mode. There have been some attempts in generating biphotons from hot atomic vapor cells, but with coherence time not exceeding 20 ns, corresponding to a bandwidth of $>$ 50 MHz that is much wider than the atomic natural linewidths they were working with \cite{GuoOE2008, GuoOE2012, RolstonOE2011}.

Here we demonstrate generating subnatural-linewidth biphotons using on-resonance SFWM in a hot $^{87}$Rb vapor cell assisted with EIT. Different from the off-resonance double-Raman scheme \cite{LvovskyPRL2012} and diamond energy-level scheme \cite{GuoOE2012, RolstonOE2011}, on which the photon bandwidth ($\sim$500 MHz) is determined by the Doppler-broadened lifetime ($\sim$2 ns) of the excited atomic states, the EIT slow-light effect can significantly prolonger the photon coherence time and narrow down the bandwidth \cite{DuJOSAB}. However directly applying the EIT-assisted SFWM scheme to a hot vapor cell, there is a serious noise problem: uncorrelated photons generated from resonance Raman scattering of the strong EIT coupling laser field overwhelms the entangled photon pairs. To overcome this problem, we coat the inner wall of the cell with paraffin to increase the atomic ground-state coherence time and apply an additional strong optical-pumping beam to eliminate the on-resonance scattering of the coupling field. This noise reduction together with other optical filtering allow us observing phase-matched biphotons with a high contrast ratio.

The experimental setup and associated atomic energy level diagram are illustrated in Fig.\ref{fig:schematic}. A paraffin-coated $^{87}$Rb (99\%, Precision Glassblowing Inc) vapor cell is placed in a temperature-stabilized hot-air heating oven, which is not shown in Fig.\ref{fig:schematic}a, and is set at 63 $^\circ$C with fluctuation less than 0.2 $^\circ$C. The length of the vapor cell is $L=0.5 $ inch and its inner diameter is $d=10$ mm. The longitudinal orientation of the cell is from east to west and there is no magnetic shielding in this experiment. The SFWM process is driven by two laser fields: The pump laser (D2 line: 780nm, $\omega_p$) is locked to the $^{85}$Rb transition $|5S_{1/2}, F=2\rangle\rightarrow|5P_{3/2}, F=3\rangle$ which is red detuned by 2.7 GHz from the $^{87}$Rb transition $|5S_{1/2}, F=1\rangle\rightarrow|5P_{3/2}, F=2\rangle$, and the coupling laser (D1 line: 795nm, $\omega_c$) is on resonance to the transition $|5S_{1/2}, F=2\rangle\rightarrow|5P_{1/2}, F=1\rangle$. The vertically polarized pump and coupling laser beams are counter propagating with the same $1/e^{2}$ beam diameter of 1.4 mm. Backward, horizontally polarized, phased-matched Stokes (780nm, $\omega_s$) and anti-Stokes (795nm, $\omega_{as}$) photon pairs are spontaneously generated, coupled into two opposing single-mode fibers (SMF), passing through optical frequency filters ($F_{s}$, and $F_{as}$), and detected by two single-photon counting modules (SPCM$_s$ and SPCM$_{as}$, Excelitas/PerkinElmer SPCM-AQRH-16-FC). The two-photon coincidence counts are recorded by a time-to-digit converter (Fast Comtec P7888) with a temporal bin width of 1 ns. Two polarization beam splitters (PBS) are used as polarization filters to distinguish the paired photons from the two driving laser beams. We find the major noise source of uncorrelated photons is from the on-resonance Raman scattering of the coupling filed following the transition $|5S_{1/2}, F=2\rangle\rightarrow|5P_{1/2}, F=1\rangle\rightarrow|5S_{1/2}, F=1\rangle$. These photons have the same central frequency and polarization as the anti-Stokes photons and can not be filtered away by the polarization and frequency filters. To clean up the residual atoms in the level $|5S_{1/2}, F=2\rangle$, we apply a strong vertically-polarized optical-pumping beam ($\omega_{op}$) that is on resonance to the transition $|5S_{1/2}, F=2\rangle\rightarrow|5P_{3/2}, F=1\rangle$. In order not to interfere the SFWM transitions, the optical-pumping beam is aligned parallel to the pump-coupling beams without overlap. The laser beam profiles on the cross section of the cell are shown in the inset of Fig.\ref{fig:schematic}a. The optical-pumping beam, with a power of 32 mW, has an $1/e^{2}$ beam diameter of 2 mm. The Stokes and anti-Stokes single-mode diameter on the cell center is 250 $\mu$m. Owning to the long ground-state coherence time because of the paraffin coating, the atoms in the level $|5S_{1/2}, F=2\rangle$ are optically pumped to the ground level $|5S_{1/2}, F=1\rangle$, an thus the Raman scattering on the anti-Stokes channel is suppressed. To further separate the generated photon pairs from the two driving laser beams, the pump and coupling laser beams are aligned with an angle of $\sim$0.1$^{\circ}$ to the Stokes and anti-Stokes directions.

\begin{figure}[!]
\includegraphics[width=8.5cm]{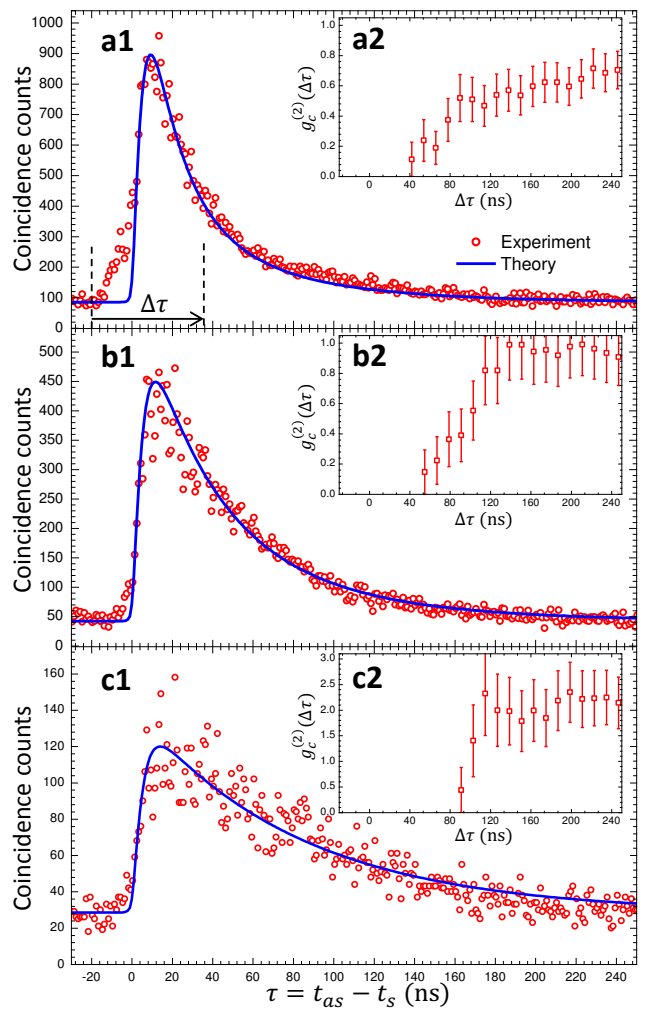}
\caption{\label{fig:Biphoton6mW} \textbf{Biphoton waveforms with controllable correlation time.} \textbf{a1-c1}, Two-photon coincidence counts, collected over 600 s with 1 ns bin width, as a function of the relative time delay $\tau$ between paired Stokes and anti-Stokes photons. The incident pumping laser power is fixed at 6 mW while the incident coupling laser power of \textbf{a1}, \textbf{b1} and \textbf{c1} are 27 mW, 9 mW and 1 mW respectively. The red circles are the experimental data. The solid blue curves are obtained numerically with the experimental parameters by taking into account the Doppler effect [\textbf{Supplementary Material}] . The corresponding measured conditional auto-correlation function $g^{(2)}_{c}$ of heralded single anti-Stokes photons are plotted in \textbf{a2-c2}.  The error bars in the data plots are the standard error which can be reduced with longer data taking time.}
\end{figure}

Figure \ref{fig:Biphoton6mW} shows the biphoton waveforms measured as two-photon coincidence counts. We fix the pump laser power at 6 mW and vary the coupling laser power, which is set to be 27, 9, and 1 mW for Fig.~\ref{fig:Biphoton6mW}\textbf{a}, \textbf{b}, and \textbf{c}, respectively. As expected, the two-photon correlation time becomes longer as we reduce the coupling laser power for narrower EIT window. Shown in Fig.~\ref{fig:Biphoton6mW}a1-c1, the biphoton waveforms are exponentially decay. The $1/e$ correlation times are 47, 60, and 94 ns, for Fig.~\ref{fig:Biphoton6mW}\textbf{a1}, \textbf{b1}, and \textbf{c1}, respectively, which all exceed the natural lifetime 25 ns of the Rb 5P excited states.  This clearly confirms the EIT effect on the SFWM biphoton generation. The blue theoretical curves are obtained numerically with the experimental parameters by taking into account the Doppler effect [\textbf{Supplementary Material}]  and agree well with the experimental data. Owning to this perfect agreement between the experiment and theory, we are able to extract the biphoton temporal wave function and joint spectrum. The bandwidths of these biphotons are 3.2, 2.6, and 1.9 MHz, for Fig.~\ref{fig:Biphoton6mW}\textbf{a1},\textbf{ b1}, and \textbf{c1}, respectively, which confirm the generated biphoton waveforms are nearly Fourier-transform limited. They are substantially narrower than the natural linewidth of 6 MHz of Rb D1/D2 lines.

To characterize the nonclassical property of the photon pair source, we confirm its violation of the Cauchy-Schwartz inequality \cite{Clauser}. Normalizing the coincidence counts to the accidental background floor in Fig.\ref{fig:Biphoton6mW}\textbf{a1-c1}, we get the normalized cross-correlation function $g^{(2)}_{s, as}(\tau)$ with maximum values of $[g^{(2)}_{s, as}]_m=11\pm1$, $11\pm2$, and $6\pm1$, respectively. With the experimental measured autocorrelations $g^{(2)}_{s, s}(0)=2.0\pm0.2$ and $g^{(2)}_{as, as}(0)=1.6\pm0.2$, we obtain the violation of the Cauchy-Schwartz inequality $[g^{(2)}_{s, as}(\tau)]^2/[g^{(2)}_{s, s}(0)g^{(2)}_{as, as}(0)]\leq1$ by factors of $38\pm8$, $38\pm11$, and $11\pm3$, for the cases in Fig.\ref{fig:Biphoton6mW}\textbf{a1}, \textbf{b1}, and \textbf{c1}, respectively.

We further quantitively verify the single-photon particle quantum nature of the heralded anti-Stokes photons by measuring its conditional auto-correlation function $g^{(2)}_c$ \cite{ConditionalG2}.  An ideal single-photon source gives $g^{(2)}_c=0$ because a single photon cannot be splitted into two. A two-photon Fock state give $g^{(2)}_c=0.5$, and and a coherent state gives $g^{(2)}_c=1$. The measured $g^{(2)}$ as a function of coincidence window width $\Delta \tau$ are plotted as Fig.\ref{fig:Biphoton6mW}\textbf{a2-c2}, which are below the two-photon threshold within their coherence time.

\begin{figure}[!]
\includegraphics[width=8.5cm]{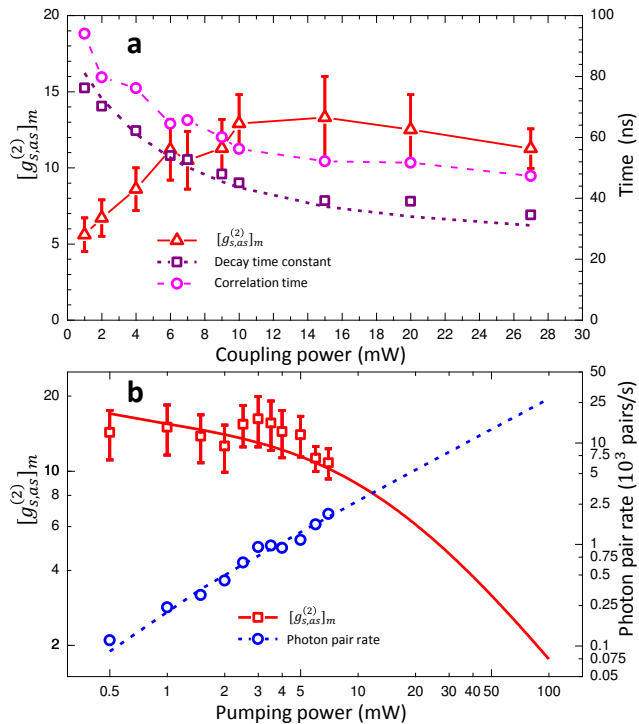}
\caption{\label{fig:cpdependent} \textbf{Biphoton generation with controllable properties.} \textbf{a}, The biphoton properties vs the coupling laser power. The pump laser power is fixed at 6 mW. The red triangular data represent the maximum of the normalized cross-correlation function, $[g^{(2)}_{s, as} ]_m$. The decay time constant (purple, square) and $1/e$ correlation time (magenta, circular) are also plotted. \textbf{b},  $[g^{(2)}_{s, as} ]_m$ and biphoton generation rate vs pump laser power. The coupling laser power is fixed at 27 mW. The solid and dashed lines are theoretical  curves.}
\end{figure}

As we reduce the coupling laser power, the EIT bandwidth becomes narrower on the anti-Stokes transition and the dispersion induced phase mismatching further constrains the Stokes-anti-Stokes two-photon joint spectrum in the SFWM process \cite{DuJOSAB}. Therefore in the EIT-assisted SFWM, the biphoton bandwidth is not determined by the lifetime of the excited states even though the photon pairs are indeed generated spontaneously. This is the mechanism to generate subnatural-linewidth biphotons from an atomic medium with broadening much wider than its natural linewidth, \textit{i.e.}, the biphotons have coherence time much longer than the natural lifetime of the atomic excited states.  Figure \ref{fig:cpdependent}\textbf{a} shows the measured decay time constant (square) $\tau_{b}$ of the biphoton waveform as a function of coupling laser power. The biphoton bandwidth can be calculated from $1/(2\pi\tau_b)$. Counting the raise time of the biphoton waveform in the rising edge around $\tau=0$, the $1/e$ correlation time is also plotted in Fig. \ref{fig:cpdependent}\textbf{a}, which is slightly longer than the decay time constant. The longest correlation time at 1mW coupling power approaches about 100 ns.  On the other side, the contrast ratio of the two-photon correlation, the maxim of the normalized cross correlation function $[g^{(2)}_{s, as}]_m$, decreases as we reduce the coupling power because of the spreading of the joint probability density in a wider time domain and the finite EIT loss at a low coupling power.  Figure \ref{fig:cpdependent}\textbf{b} shows the $[g^{(2)}_{s, as}]_m$ and photon pair generation rate as a function of the pump power. The fiber coupling efficiencies, filter transmissions, SPCM quantum efficiencies have been all taken into account to obtain the photon-pair generation rate. While the photon pair rate is proportional to the pump laser power, the  $[g^{(2)}_{s, as}]_m$ reduces at a high pump power. limited by our maximum available pump laser power of 7 mW, we produced about 2000 pairs/s. The solid curve is a theoretical plot obtained by taking into account the uncorrelated noise photons. The nearly constant $[g^{(2)}_{s, as}]_m$ at a low pump power suggest that the accidental coincidence counts dominate the flat background floor of the two-photon correlation. As the threshold of the $[g^{(2)}_{s, as}]_m$ is 2.0 for violating the Cauchy-Schwartz inequality, the nonclassical property of the photon source is still preserved at 90 mW, which corresponds to a generation rate of about 25,000  pairs/s.

In summary, we experimentally demonstrate generation of subnatural-linewidth biphotons from a hot paraffin-coated $^{87}$Rb vapor cell using EIT-assisted SFWM.  The biphoton coherence time, which is controlled by varying the coupling laser power, can be as long as 94 ns, which corresponds to a spectrum bandwidth of 1.9 MHz that is substantially narrower than the natural linewidth 6 MHz of Rb D1/D2 transitions. Their quantum nature is verified by violating the Cauchy-Schwartz inequality and measuring the conditional autocorrelation function. It can be used to generate nearly-pure heralded single photons \cite{DuPRA2015}. Its Lorentzian line shape and exponential waveform with tunable time constant is perfect for interacting with atoms \cite{DuPRL2012} and coupling to an optical cavity \cite{DuPRL2014}. As compared to the cold-atom-based narrowband biphoton source, the hot atomic vapor cell configuration is much simpler and easier for operation and maintenance. Our demonstration can lead to miniature narrowband biphoton source that is suitable for practical quantum applications.


\vspace{0.5cm} \noindent\textbf{Acknowledgements} \noindent The authors thank K. Zhao at Fudan University for helpful discussions. T.K.A. C. acknowledges support from the Undergraduate Research Opportunities Program at the Hong Kong University of Science and Technology. The work was supported by Hong Kong Research Grants Council (Project No. 16301214).

\vspace{0.5cm} \noindent\textbf{Author Contributions} \noindent S.D. conceived of the idea. C.S. and S.D. designed the experiment. C.S., P.C., T.K.A.C. and L.Z. performed the experiment. C.S. analyzed the data. S.D., C.S., and Y.Xiao discussed the feasibility of the experiment. S.D. and M.M.T.L. directed the project. All authors contributed to the final manuscript.

\vspace{0.5cm} \noindent\textbf{Author Information} \noindent The authors declare no competing financial interests. Correspondence and requests for materials should be addressed to S.D. (dusw@ust.hk).

\end{document}